\documentclass[aps, prl, 10pt, twocolumn,preprintnumbers,amsmath,amssymb,superscriptaddress,showkeys,floatfix]{revtex4-1}
\usepackage{amssymb}
\usepackage{amsmath}
\usepackage{graphicx}
\usepackage{dsfont}
\usepackage{color}

\newcommand{\kbf}{\mathbf{k}}

\begin{document}
\title{Tunable Flux Vortices in 2D Dirac Superconductors}

\author{Sina Zeytino\u{g}lu}
\affiliation{Institute for Theoretical Physics, ETH Z\"urich, CH-8093 Zurich, Switzerland.}
\affiliation{Institute for Quantum Electronics, ETH Z\"urich, CH-8093 Zurich, Switzerland.}

\author{Atac \.Imamo\u{g}lu}
\affiliation{Institute for Quantum Electronics, ETH Z\"urich, CH-8093 Zurich, Switzerland.}

\author{Sebastian Huber}
\affiliation{Institute for Theoretical Physics, ETH Z\"urich, CH-8093 Zurich, Switzerland.}

%

\date{\today }

\begin{abstract}
The non-trivial geometry encoded in the Quantum Mechanical wavefunctions has important consequences 
for its single-particle as well as many-body dynamics. Yet, our understanding of how the geometry of the single-particle
eigenstates are manifest in the characteristics of a many-particle system is still incomplete.
Here, we demonstrate how the single-particle Berry curvature modifies the fluxoid quantization of a 
two dimensional Bardeen-Cooper-Schrieffer (BCS) superconductor, and discuss the experimental scenarios where this anomalous
quantization is expected to be realized. 
\end{abstract}

\pacs{}
\maketitle

The realization that the Hilbert space of a quantum mechanical system can be endowed with a nontrivial
geometry resulted in the discovery of a plethora of novel phenomena both in single- and many-particle
quantum systems. The geometric considerations associated with single-particle wavefunctions have
led to the discovery and classification of non-interacting topological insulators (TI's) \cite{Schnyder2009,Kitaev2009,Fu2011,Bradlyn2017}, 
as well as to the investigation of the novel properties associated with defects in these phases \cite{Ivanov2001, Teo2017}.
On the other hand, the fascinating physics of interacting particles occupying non-trivial single-particle states
was investigated in the framework of fractional quantum Hall systems (FQHS's) \cite{Laughlin1983,Girvin1999}
 and fraction Chern Insulators (FCI's) \cite{Regnault2011,Neupert2011}. However, the strongly correlated nature of 
 these systems makes it difficult to establish a direct relation between the role of non-trivial geometry on the 
 single- and the many-particle level.
  
In this Letter, we demonstrate a clear connection between the non-trivial geometry of a two-dimensional (2D) single-particle 
band structure and the response properties of a weakly interacting system. 
To this end, we consider a Bardeen-Cooper-Schrieffer (BCS) superconductor \cite{BCS1957},
consisting of Cooper pairs of electrons whose single-particle
states carry non-trivial geometry. We show that in a 2D system, the non-trivial Berry curvature of the single-particle states 
modifies the conventional fluxoid quantization associated with the superconducting order parameter,
resulting in fractional flux vortices as well as in an unconventional Josephson response.


%
%
%
%

\begin{figure}[htbp]
\begin{center}
\includegraphics[width=0.4\textwidth]{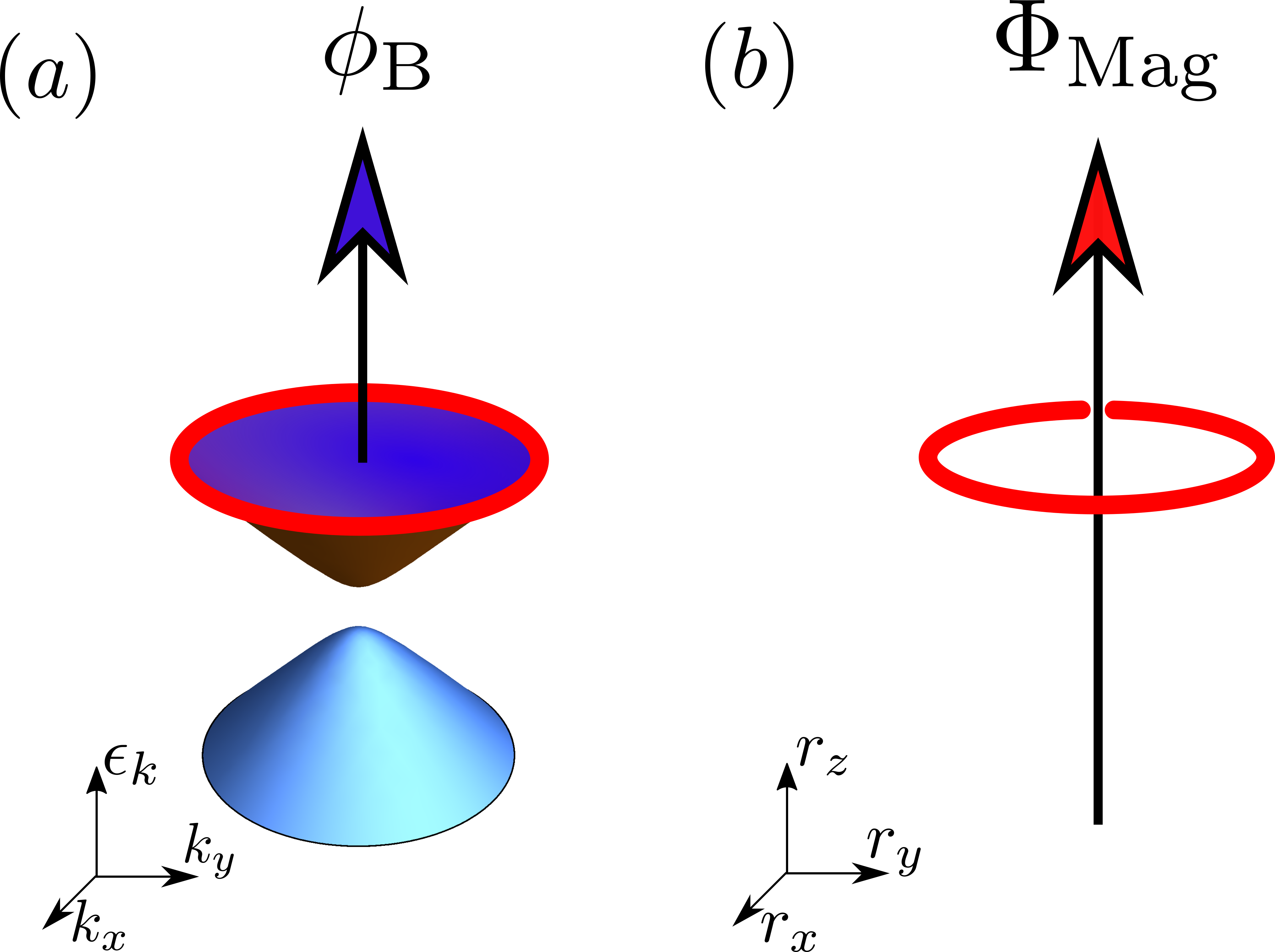}
\caption{(a) The fractional angular momentum of single-particles on a circular Fermi surface enclosing non-zero 
Berry flux $\phi_{\mathrm{B}}$. Similar to the magnetic field in real space, the Berry curvature carries an intrinsic angular momentum
which is acquired by the single-particle states on the Fermi surface. The Berry flux $\phi_{\mathrm{B}}$ enclosed in the Fermi surface determines the fractional part of the angular momentum associated with the states on the Fermi surface. (b) The fractional
angular momentum due to the Berry flux can be considered as the momentum space analog of the anyon construction
in Ref. \cite{Wilczek1982a, Wilczek1982b}.}
\label{fig:Analogy}
\end{center}
\end{figure}
\begin{figure}[htbp]
\begin{center}
\includegraphics[width=0.5\textwidth]{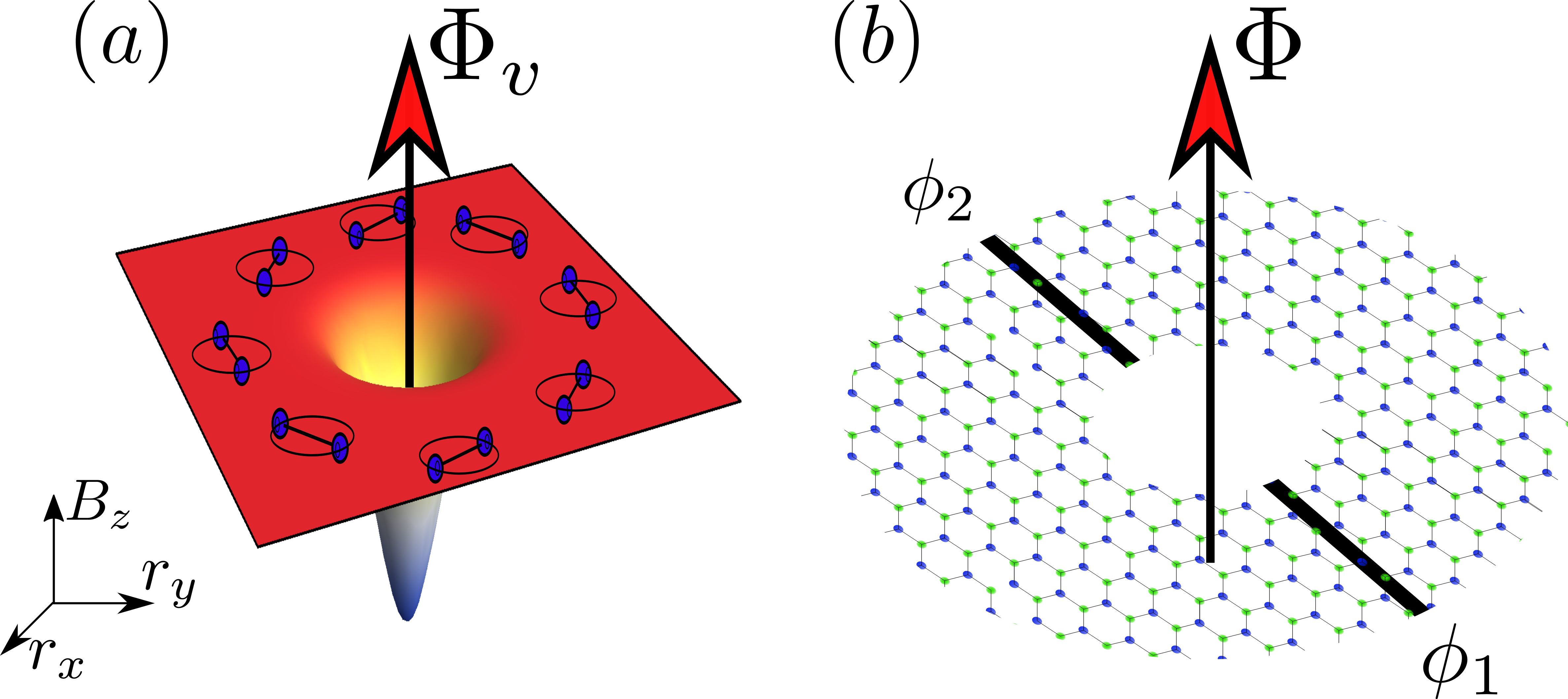}
\caption{The illustration of the configuration of the superconducting order parameter which allows 
stabilization of a vortex with fractional magnetic flux quantum $\Phi_0 \left(\frac{\phi_{\mathrm{B}}}{\pi}\right)$.
The single-valuedness of the order parameter is satisfied only is the relative coordinate of the Cooper
pairs rotate by a multiple $q$ of $2\pi$ around the vortex. Such a configuration realizes an adiabatic 
evolution of the order parameter as a function of the center of mass position. The Berry flux acquired
by the order parameter contributes to the overall phase of the superconductor, modifying the conventional fluxoid quantization.
(b) The modification of the fluxoid quantization also allows for building SQUID's with tunable periodicity.}
\label{fig:SQUID}
\end{center}
\end{figure}

The fluxoid quantization originates from the fact that deep inside a bulk superconductor, 
the overall phase of the order parameter is constant. 
In conventional superconductors, this overall phase has contributions from two sources: 
the external electromagnetic vector potential  and the $U(1)$ phase of the order parameter.
Hence, the fluxoid quantization can be understood as the total screening of the electromagnetic
vector potential by the phase of the order parameter field \cite{Tinkham2004}. However, as we show below, an additional contribution
to the overall phase emerges for a BCS superconductor whose normal state Fermi surface encloses a non-trivial Berry flux
$\phi_{\mathrm{B}}$ [see Fig.~$\ref{fig:Analogy}$ (a)]. This additional contribution is related to the orbital currents in the superconductor and
allows for screening of the electromagnetic vector potential by the Aharanov-Bohm phase associated with the internal rotation of the Cooper pair. In particular,  
a $2\pi$ rotation of the relative coordinate of the Cooper pair results in a change in the overall phase by $2\phi_{\mathrm{B}}$,
allowing stable fractional flux vortices [see Fig.~$\ref{fig:SQUID}$ (a)] .
  

We consider the simplest model for a 2D normal state which has a rotationally symmetric dispersion
and a non-trivial Berry flux enclosed by the Fermi surface. The Hamiltonian for the massive Dirac model is
\begin{align}
H_0(k) =\left( f^{\dagger}_{1,\mathbf{k}} \quad f^{\dagger}_{2,\mathbf{k}} \right) \left[\mathbf{d}(\mathbf{k})\cdot \boldsymbol{\sigma} - \mu \mathds{1}\right]  \left(\begin{array}{c} f_{1,\mathbf{k}} \\ f_{2,\mathbf{k}} \end{array} \right),
\label{eq:Dirac}
\end{align}
where $\boldsymbol{\sigma}= (\sigma_x,\sigma_y,\sigma_z)$ is the vector of Pauli matrices, $\mathds{1}$ is the identity matrix,
and $f^{\dagger}_{i,\mathbf{k}}$ $(f_{i,\mathbf{k}})$ are the creation (annihilation) operators for particles with 
momentum $\mathbf{k}$ and spin or pseudo-spin label $i$.
In the following, we let $|c(k)\rangle$ and $|v(k)\rangle$ denote the eigenvectors of $H_0(k)$,
which correspond to conduction and valence bands, respectively.
The single-particle Hamiltonian is parametrized by (i) $\mathbf{d}(\mathbf{k}) = ( v_{\mathrm{F}} k_x, v_{\mathrm{F}} k_y,\delta)$, 
with $2 \delta$ characterizing the minimum energy gap between the conduction and valence bands, (ii) the Fermi 
velocity $v_{\mathrm{F}}$, which is taken to be unity for the rest of the Letter, and (iii) the chemical potential $\mu$, which
we set to lie in the conduction band. We note that the single-particle Hamiltonian in Eq. ($\ref{eq:Dirac}$) is a good effective
low-energy description a for family of real materials, most notably transitional metal
dichalcogenide (TMD) monolayers near the high symmetry $K$ points \cite{Wang2012,Xu2014}, or surfaces of
three-dimensional (3D) TI's in the proximity of a ferromagnetic insulator \cite{Chen2010}. We give a more 
comprehensive list of experimental scenarios which can be modelled by Eq. ($\ref{eq:Dirac}$) in our conclusion. 

The massive Dirac model in Eq.~($\ref{eq:Dirac}$) exhibits non-trivial geometry for both the conduction and valence
band. The non-trivial geometry can be understood as a result of the the Berry connection at momentum $\mathbf{k}$
\begin{align}
\mathbf{A}_{\mathrm{c}}(\mathbf{k})= i \langle c(\mathbf{k}) |\nabla_{\mathbf{k}} c(\mathbf{k}) \rangle =\frac{1}{k} \sin{(\theta_k/2)}^2 \mathbf{e}_{\phi},
\label{eq:BerryConn}
\end{align}
where $\theta_k \equiv \arccos{\frac{\delta}{\sqrt{\delta^2 + k^2}}}$ and $\mathbf{e}_{\phi}$ is the unit vector in the azimuthal 
direction. Most importantly for this work, the non-trivial connection results in a non-zero Berry flux $\phi_{\mathrm{B}}$
enclosed in the conduction band Fermi surface
\begin{align}
\phi_{\mathrm{B}}(k_{\mathrm{F}}) = \oint_{\mathrm{FS}} dk \cdot \mathbf{A}_{\mathrm{c}}(\mathbf{k}) = \pi \left(1-\frac{\delta}{\sqrt{\delta^2 + {k_{\mathrm{F}}}^2}}\right),
\label{eq:BerryPhase}
\end{align}
where $k_{\mathrm{F}}$ denotes the Fermi momentum.

The normal state described by the Eq.~($\ref{eq:Dirac}$) has a superconducting 
instability with respect to attractive interactions between electrons.
In general, the interaction between the electrons can be written in second quantization in the (pseudo-) spin basis ($m,n \in \{1,2\}$) as  
\begin{align}
H_{int} = \sum_{\kbf_1,\kbf_2,\kbf_3,\kbf_4, m,n} V_{\kbf_1,\kbf_2,\kbf_3,\kbf_4}^{n,m} f_{m,\kbf_1}^{\dagger} f_{n,\kbf_2}^{\dagger} f_{n,\kbf_3} f_{m,\kbf_4},
\label{eq:IntHam}
\end{align}
where  $V_{\kbf_1,\kbf_2,\kbf_3,\kbf_4}^{n,m}<0$ for all $\{\kbf_{i}\}$, and $V_{\kbf_1,\kbf_2,\kbf_3,\kbf_4}^{n,m} \propto \delta_{\kbf_1+\kbf_2,\kbf_3+\kbf_4}$ ensures momentum conservation. 

For BCS superconductivity, the only relevant degrees of freedom are th zero momentum Cooper pairs consisting
electrons on the conduction band Fermi surface \cite{BCS1957}. Hence, we consider only the terms where
$\kbf_1=-\kbf_2$ and $\kbf_3 = - \kbf_4$ in Eq.~($\ref{eq:IntHam}$).
Projecting the interaction Hamiltonian onto the conduction band, we obtain
\begin{align}
\tilde{H}_{int} =  \sum_{\kbf,\kbf' } U_{\kbf,\kbf'} c_{\kbf}^{\dagger} c_{-\kbf}^{\dagger} c_{\kbf'} c_{-\kbf'},
\label{eq:ProjInt}
\end{align}
where $c^{\dagger}_{\kbf}|0\rangle  \equiv \left(\alpha_{c,\kbf}^{(1)} f_{1,\kbf} + \alpha_{c,\kbf}^{(2)} f_{2,\kbf}\right)^{\dagger}|0\rangle= |c(\mathbf{k})\rangle$.
The strength of interactions projected to the conduction band is  
\begin{align}
U_{\kbf,\kbf'} = \sum_{n,m}  V_{\kbf,-\kbf,\kbf',-\kbf'}^{n,m} \alpha^{(n) *}_{c \kbf} \alpha^{(m)*}_{c -\kbf} \alpha^{(m)}_{c \kbf'}\alpha^{(n)}_{c -\kbf'},
\end{align}
here $\phi_k$ is the azimuthal angle of the $\mathbf{k}$. 

In the context of our work, the most remarkable feature of the projected interaction in Eq. ($\ref{eq:ProjInt}$) is that when we 
restrict the momenta on the Fermi surface (i.e., $\kbf=k_F \hat{\mathbf{k}}$), the coefficients $U_{k_{\mathrm{F}}}^{(l)}$
of the partial-wave expansion, defined through 
$U_{k_{\mathrm{F}}\hat{\mathbf{k}},k_{\mathrm{F}}\hat{\mathbf{k'}}} = \sum_{l}U_{k_{\mathrm{F}}}^{(l)} e^{i(\phi_{k_{F}}-\phi_{k_{F}'})}$,
are redistributed with respect to the coefficients $V^{(l)}_{k_{\mathrm{F}}}$ of the bare interaction.
For example, in the massive Dirac model in Eq. ($\ref{eq:Dirac}$), and $s$-wave interactions in the orbital basis
$V^{(l)}\propto \delta_{l,0}$, the partial-wave coefficients of the interaction projected onto the conduction band is purely $p$-wave.
In particular,  $U_{k_{\mathrm{F}}}^{(l)}\propto \delta_{l,1}$, with
\begin{align}
 U_{k_{\mathrm{F}}}^{(1)} &= \left(V_{k_{\mathrm{F}}}^{12}\right)^{(0)} \left(1-\frac{\phi_B}{2\pi}\right)\frac{\phi_B}{2\pi},
 \label{eq:AngMomComp}
\end{align}
where we used the constraints $(V_{k_{\mathrm{F}}}^{nn})^{(l)}=0$ and $U_{k_{\mathrm{F}}}^{(l)} = 0$ for even $l$ arising from the
fermionic anti-commutation relations obeyed by operators  $f_{m,k},\, f^{\dagger}_{m,k}$ and $c_{k},\, c^{\dagger}_{k}$, respectively.
In general, when the bare interaction potential has multiple partial wave components, the coefficients $U^{(l)}_{k_{\mathrm{R}}}$ obey
\begin{align}
\nonumber U^{(l)}_{k_{\mathrm{F}}} &=  \left(V_{k_{\mathrm{F}}}^{11}\right)^{(l)}\left(1-\frac{\phi_B}{2\pi}\right)^2 + \left(V_{k_{\mathrm{F}}}^{12}\right)^{(l-1)} \left(1-\frac{\phi_B}{2\pi}\right)\frac{\phi_B}{2\pi} \\
&+ \left(V_{k_{\mathrm{F}}}^{22}\right)^{(l-2)}  \frac{\phi_B}{2\pi}^2.
\end{align}
The redistribution of the partial-wave components can be understood as a direct consequence of the non-trivial Berry phase inclosed in the 
single-particle Fermi surface. Eq.~($\ref{eq:AngMomComp}$) expresses an effect that is similar to the emergence of effective 
$p$-wave pairing from $s$-wave interactions, as studied in TI surfaces \cite{Fu2008} proximity coupled to an $s$-wave superconductor, 
as well as in the spectroscopic \cite{Srivastava2015b,Zhou2015} and transport \cite{Yao2008} signatures
of Berry curvature in 2D excitons. Furthermore, in three dimensions, the above discussion 
leads to the prediction of nodal superconductivity in Weyl semimetals \cite{Murakami2003,Li2018}.


Next, we treat the superconducting phase where the Cooper pairs consist of electrons on the 
conduction band Fermi surface which encloses a Berry flux of $\phi_{B}(k_{\mathrm{F}})$.
In the weak coupling regime, we can describe the superconducting ground state using the BCS ansatz
 \begin{align}
 |\mathrm{BCS}\rangle =\prod_{k} \left(u_\mathbf{k} + v_\mathbf{k} c^{\dagger}_{\mathbf{k}}c^{\dagger}_{-\mathbf{k}}\right) |0\rangle,
 \label{eq:BCSWF}
 \end{align}
 where $|0\rangle$ is the electronic vacuum of the conduction band. The BCS ansatz parametrizes the expectation value of the 
 Cooper pair $\langle \mathrm{BCS} |c_{\mathbf{k}} c_{-\mathbf{k}}|\mathrm{BCS}\rangle = u_{\mathbf{k}}^*v_{\mathbf{k}} \equiv F_{\mathbf{k}}$.
 Given the BCS gap equation $\Delta_{\mathbf{k}} = \sum_{\mathbf{k}'} U_{\mathbf{k}\mathbf{k}'} F_{\mathbf{k}'} $, the expectation value $F_{\mathbf{k}}$ can be 
 written as $|F_{\mathbf{k}}| e^{i l_0 \phi_{\mathbf{k}_F}}$, where $l_0$ maximizes $|U_{\mathbf{k}_{F},\mathbf{k}_{F}'}^{l_0}|$ under the 
 constraint that the Cooper pair wavefunction in a given (pseudo-) spin representation is anti-symmetric. 
 
 Naively, one could conclude from the above discussion that the expectation value of the many-body angular momentum operator
 is $\hbar l_0 n_c$, where we defined the number of condensed pairs as $n_c~\equiv~\sum_{\mathbf{k}} F_{\mathbf{k}}$.
 However, a more careful calculation shows that this conclusion is incorrect in the presence of the non-trivial Berry curvature.
The expectation value of the relative angular momentum operator is (see Supplementary Material)
 \begin{align}
 \nonumber & \langle L_{\mathrm{rel}}\rangle \equiv - \frac{1}{2} \langle \mathrm{BCS}|\sum_{i \neq j} (\hat{r}_i - \hat{r}_j) \times (\hat{p}_i - \hat{p}_j)|\mathrm{BCS}\rangle \\
 \nonumber &=   \hbar \sum_{\mathbf{k}} F^*(\mathbf{k})   \left(- i \partial_{\phi}\right)  F(\mathbf{k}) + \sum_{\mathbf{k}}F^{*}(\mathbf{k}) \hbar \frac{\phi_{\mathrm{B}}(k_{\mathrm{F}})}{\pi} F(\mathbf{k})\\
 &=  n_c \hbar \left(l_0  +  \frac{\phi_{B}(k_{\mathrm{F}})}{\pi} \right).
 \label{eq:AngMomExp}
 \end{align}
We observe that the expectation value of the angular momentum operator is shifted with respect
to the naive guess above by twice the angular momentum associated with $\phi_{\mathrm{B}}(k_{\mathrm{F}})$ in
Eq.~($\ref{eq:BerryPhase}$). Importantly, the Berry flux contribution can be traced back to the 
 fact that the second quantized position operator projected onto the conduction band is \cite{Blount1962,Girvin1999}
 \begin{align}
 \nonumber \hat{r} &= \sum_{\mathbf{k},\mathbf{k}'}\langle c(\mathbf{k}') |\hat{r}| c(\mathbf{k}) \rangle c_{\mathbf{k}'}^{\dagger}c_{\mathbf{k}}\\
 & = \sum_{\mathbf{k},\mathbf{k}'}\left[ -i \partial_{\mathbf{k}}  + A_c(\mathbf{k}) \right]  \delta_{\mathbf{k},\mathbf{k}'} c_{\mathbf{k}'}^{\dagger}c_{\mathbf{k}}.
\end{align} 
Equivalently, the angular momentum of the Cooper pair is given by the eigenvalue of the covariant derivative 
in momentum space, $D_{\phi} = -i\partial_{\phi} + \frac{2\phi_{\mathrm{B}}(k_{\mathrm{F}})}{2\pi}$. 

This result can be intuitively understood by considering the analogy between the single-particle 
wavefunctions on the Fermi surface enclosing the Berry flux $\phi_{\mathrm{B}}$, and the wavefunctions
of a particle on a ring which encloses a magnetic flux $\Phi_{\mathrm{Mag}}$ (see Fig. $\ref{fig:Analogy}$). As discussed 
in Ref. \cite{Wilczek1982a,Wilczek1982b}, in both cases, the eigenfunctions have angular momenta
whose values are shifted with respect to their conventional integer values by a fraction associated with $\phi_{\mathrm{B}}$ or 
$\Phi_{\mathrm{Mag}}$. Thus, the unconventional expectation value of the Cooper pair angular momentum
can be understood as the sum of the shifted angular momenta of their constituent particles.

What are the physical effects of the fractional angular momentum expectation value of the Cooper 
pairs? We show that one interesting consequence of the fractional angular momentum 
is its effect on the orbital currents. 
We calculate the orbital current contribution in the Ginzburg Landau (GL) free energy from the microscopic model 
using the formalism of functional integrals (Supplementary Material) \cite{Altland2010,Mudry2014}. 
To this end, we consider a configuration of the order parameter where the relative orientation
of the Cooper pairs is adiabatically rotated as a function of their center of mass position $R$ [see Fig.~$\ref{fig:SQUID}$ (a)]. 
As a result, the gradient term in the GL free energy density can be shown to be
\begin{align}
\nonumber &F_{\mathrm{grad}} = \frac{\hbar^2 \rho_s}{2m^*} \bigg( -2\frac{e}{\hbar} \mathbf{A}_{\mathrm{e.m.}}(R)\\
&+ \nabla  \theta(R)+\left( l_0 +\frac{\phi_{\mathrm{B}}}{\pi} \right)\nabla \phi_k(R)\bigg)^2, 
\label{eq:GradEnergy}
\end{align}
where $\rho_s \equiv |\Delta|^2$ is the superfluid density for an isotropic superconductor, $m^*$ is the mass of the Cooper pair,
$\mathbf{A}_{\mathrm{e.m.}}(R)$ is the electromagnetic vector potential, and $\theta(R)$ is the 
$U(1)$ phase of the order parameter. We omit the $k_{\mathrm{F}}$ dependence of the Berry flux for simplicity.
Taking the variation of the free energy with respect to $\mathbf{A}_{\mathrm{e.m.}}$, we find that the supercurrent
including the effect of a variation in the relative orientation of the Cooper pair is
\begin{align}
\nonumber \mathbf{J}_s &= \frac{\delta F}{\delta \mathbf{A}_{\mathrm{e.m.}}} = \frac{\hbar^2 \rho_s}{m^*} \bigg[ -2\frac{e}{\hbar} \mathbf{A}_{\mathrm{e.m.}}(R)\\
&+ \nabla \theta(R)+(l_0+ \phi_{\mathrm{B}}/\pi) \nabla \phi_k(R) \bigg].
\label{eq:SuperCurr}
\end{align} 
While the first and the second term in the above expression can be attributed to the charge and 
mass currents in a conventional superconductor, respectively, the third term contains the effect of the Berry flux 
enclosed in the Fermi surface. We note that because the normal state has a rotationally symmetric (isotropic) Fermi surface,
the energy associated with the orbital currents goes to zero for configurations where the 
wavelength of the variation is taken to infinity. 

The form of the orbital current contribution in Eq.~($\ref{eq:SuperCurr}$), modifies the possible vortex states of the superconductor.
As is well known, the magnetic field inside a superconductor is screened due to the Meissner effect~\cite{Orlando1991, Altland2010},
resulting in the condition that away from the vortex core, the supercurrents should vanish.
This statement is captured by the following equation:
\begin{align}
\oint_{C} d\mathbf{l} \cdot \mathbf{J}_s = 0,
\label{eq:Quant}
\end{align}
where $C$ denotes a path around the vortex core, and we assumed that
the distance between any point on $C$ and the vortex core is much larger than the 
effective London penetration length \cite{Pearl1964,Fetter1980} of the 2D superconductor. Inserting $
(\ref{eq:SuperCurr})$ into ($\ref{eq:Quant}$), we find the minimum value of the 
magnetic flux $\Phi_v \equiv \oint_{C} dl \cdot  \mathbf{A}_{\mathrm{e.m.}} $ the vortex can accommodate to be
\begin{align}
\Phi_v = \frac{2\pi \hbar}{2e} \left[p +(l_0+\phi_{\mathrm{B}}/\pi) q \right],
\label{eq:flux}
\end{align}
where  $p,q \in \mathds{Z}$ denote the winding numbers of the $U(1)$ phase and the orientation angle $\phi_{\mathbf{k}}$
of the Cooper pairs along the contour, respectively. 
The minimal magnetic flux $\Phi_{\mathrm{min}}$ enclosed in the vortex core in a superconductor on a band with 
non-trivial Berry curvature is given by setting $q=1$ and $p = -l_0$:
\begin{align}
\Phi_{\mathrm{min}} = \frac{h}{2e} \phi_{\mathrm{B}}/\pi.
\end{align}
Given that $2 \phi_{\mathrm{B}}$ can be interpreted as the Aharanov-Bohm phase acquired by the Cooper pair wavefunction as the orientation
of the pair is adiabatically rotated by $2\pi$, the configuration with $q=1$ corresponds to a vortex with fractional flux quantum $\Phi_{min}$
that is stable due the screening of the electromagnetic field by the Aharanov-Bohm phase  [see Fig. $\ref{fig:SQUID}$ (a)].

The modification of the flux quantization in the presence of non-trivial Berry curvature and an isotropic Fermi surface also bears interesting
consequences for a superconducting quantum interference device (SQUID)  [see Fig.~$\ref{fig:SQUID}$ (b)].
The total current $i_T$ through the SQUID consisting of two Josephson junctions is 
\begin{align}
i_T = i_1+i_2 = 2 I_c \cos{\left(\frac{\phi_1-\phi_2}{2}\right)}\sin{\left(\frac{\phi_1+\phi_2}{2}\right)},
\label{eq:SQUID}
\end{align}
where $I_c$ is the critical current of each junction and $\phi_j$ denotes the phase difference 
between the two sides of the $j^{th}$ Josephson junction. When the thickness of the superconducting regions
connecting the two Josephson junctions is much larger then the penetration length, we can define a contour 
$C$ for which Eq.~($\ref{eq:Quant}$) is satisfied. Then the relation between the 
phase difference $\phi_1-\phi_2$ and the total flux
$\Phi$ is  
\begin{align}
\frac{\Phi}{\Phi_0}  = \frac{\phi_1-\phi_2}{2\pi}  + p + (l_0 + \phi_{\mathrm{B}}/\pi) q,
\label{eq:SQUIDrel}
\end{align}
where the winding numbers $p,q \in \mathds{Z}$ have the same meaning as in 
Eq. ($\ref{eq:flux}$).
Inserting ($\ref{eq:SQUIDrel}$) into ($\ref{eq:SQUID}$), the total current simplifies to
\begin{align}
i_T = 2 I_c \cos{\left( \phi_{\mathrm{B}} q + \pi \frac{\Phi}{\Phi_0}\right)}\sin{\left(\phi_1+ \phi_{\mathrm{B}} q+ \pi \frac{\Phi}{\Phi_0}\right)}.
\end{align}
Maximizing the driving current, we obtain
\begin{align}
i_{\mathrm{max}} \approx 2I_c \left| \cos{\left( \phi_{\mathrm{B}}q+ \pi \frac{\Phi_{\mathrm{ext}}}{\Phi_0}\right)}\right|.
\label{eq:SQUIDmax}
\end{align}
where we neglected the self-inductance of the superconducting loop 
(i.e., $\Phi = \Phi_{\mathrm{ext}}$). The value of the winding number $q$ in Eq.~($\ref{eq:SQUIDmax}$) 
is given by the configuration which minimizes the kinetic energy of the system for a given $\Phi_{\mathrm{ext}}$. As a result, the 
Josephson current through the SQUID has a periodicity which is a fraction of the conventional SQUID.

Before we conclude, we discuss possible experimental realizations of the physics presented above. 
The experimental setups where the single-particle Hamiltonian
in Eq.~($\ref{eq:Dirac}$) can be realized include (i) surface modes of a 3D TI  in the \textcolor{black}{proximity of a ferromagnetic insulator} \cite{Fu2008,Chen2010}
(ii) an isotropic anomalous Hall system  \cite{Nagaosa2010} (iii) a single $K$ - valley of a TMD monolayer 
\cite{Xu2014, Srivastava2015a} and (iv) a single $K$ - valley of a bilayer graphene biased by an in plane 
electric field \cite{Oostinga2008}. 

The effects discussed above will also be present if each electron comprising the Cooper pair
can be described by a separate massive Dirac Hamiltonian with the same Berry flux enclosed by each Fermi surface. 
Such systems include (v)  a graphene monolayer on a hexagonal anti-ferromagnetic substrate \cite{Qiao2014} 
and (vi) a cold atom realization of a weakly-doped Chern insulator \cite{Aidelsburger2015,Jotzu2014}.

For systems (i), (ii), and (v) above, the effects discussed above can be in principle observed by inducing
a superconducting gap via proximity effect to a conventional $s$-wave superconductor.
However, this strategy is not viable for systems which do not break time-reversal (TR) symmetry [(iii) and (iv)]
since proximity effect only allows pairing between zero total momentum Cooper pairs, and the $K$ and $-K$
valleys in (iii) and (iv) carry opposite Berry curvature.
Consequently, the superconducting gap in (iii) and (iv) should be opened via intrinsic attractive 
\cite{ Rahimi2017} or repulsive \cite{Hsu2017} interactions which 
favor intravalley pairing. Lastly, in cold atom settings (vi), reliable Feshbach resonances allow greater freedom to 
realize pairing with desired properties.

In conclusion, we have demonstrated a direct relation between a non-trivial single-particle Berry phase and 
the modification of the fluxoid quantization in a BCS superconductor. A natural extension of our treatment is
to include the effects of anisotropy and disorder.
We also emphasize that our approach to combine
the projected many-body position operators and with a mean field ansatz can be applied to other weakly correlated systems
consisting of excitons \cite{Yao2008} or polarons \cite{Sidler2016}.

We acknowledge useful discussions with Manfred Sigrist, Dima Geshkenbein, Aline Ramires, Ovidiu Cotlet, Mark Fisher, and Tomas Bzdusek , Jose Luis Lado. This work is supported by a European Research Council (ERC) Advanced investigator grant (POLTDES) and Consolidator grant (TopMechMat).

\end{document}